\documentclass[12pt,headings=big,numbers=noenddot,DIV=14]{scrartcl}
\usepackage{graphicx,epsfig}
\usepackage{amsmath,amssymb}
\usepackage{array}
\usepackage{arydshln}
\usepackage{color}
\usepackage[nosort]{cite}
\usepackage{enumerate}
\usepackage{subfigure}
\usepackage{bbm}
\usepackage{slashed}
\usepackage{multirow}
\usepackage{fancybox}
\usepackage[bookmarks=false]{hyperref}
\addtokomafont{disposition}{\rmfamily\boldmath}

\newcommand{\be}{\begin{equation}}
\newcommand{\ee}{\end{equation}}
\newcommand{\bea}{\begin{eqnarray}}
\newcommand{\eea}{\end{eqnarray}}
\newcommand{\beq}{\begin{equation}}
\newcommand{\eeq}{\end{equation}}
\newcommand{\beqa}{\begin{eqnarray}}
\newcommand{\eeqa}{\end{eqnarray}}
\newcommand{\ba}{\begin{array}}
\newcommand{\ea}{\end{array}}

\definecolor{nicered}{rgb}{0.7,0.1,0.1}
\newcommand{\V}{\boldsymbol{V}}

\newcommand{\Vu}{\boldsymbol{V_u}}
\newcommand{\Vd}{\boldsymbol{V_d}}

\def\I{\mathcal{I}}

\newcommand{\qL}{\boldsymbol{q}_{\boldsymbol{L}}}
\newcommand{\uR}{\boldsymbol{u}_{\boldsymbol{R}}}
\newcommand{\dR}{\boldsymbol{d}_{\boldsymbol{R}}}

\newcommand{\qLbar}{\boldsymbol{\bar q}_{\boldsymbol{L}}}
\newcommand{\uRbar}{\boldsymbol{\bar u}_{\boldsymbol{R}}}

\DeclareFontFamily{OT1}{pzc}{}
\DeclareFontShape{OT1}{pzc}{m}{it}{<-> s * [1.10] pzcmi7t}{}
\DeclareMathAlphabet{\mathpzc}{OT1}{pzc}{m}{it}

\title{\Large\bf\boldmath
Less Minimal Flavour Violation
}
\date{\small{E-mail: \href{mailto:barbieri@sns.it}{barbieri@sns.it}, \href{mailto:dario.buttazzo@sns.it}{dario.buttazzo@sns.it}, \href{mailto:filippo.sala@sns.it}{filippo.sala@sns.it}, \href{mailto:david.straub@sns.it}{david.straub@sns.it}}}
\author{\normalsize Riccardo Barbieri, Dario Buttazzo, Filippo Sala, and David M. Straub
\\{\em\normalsize Scuola Normale Superiore and INFN, Piazza dei Cavalieri 7, 56126 Pisa, Italy}
}

\begin{document}

\maketitle
\setcounter{tocdepth}{2}

\begin{abstract}
\noindent
We consider the approximate $U(2)^3$ flavour symmetry exhibited by the quark sector of the Standard Model and all its possible breaking terms appearing in the quark Yukawa couplings. Taking an Effective Field Theory point of view, we determine the current bounds on these parameters, assumed to control the breaking of flavour in a generic extension of the Standard Model at a reference scale $\Lambda$. In particular, a significant bound from $\epsilon^\prime/\epsilon$ is derived, which is relevant to Minimal Flavour Violation as well.
In the up-quark sector, the recently observed CP violation in $D\rightarrow \pi^+ \pi^-$, $K^+ K^-$ decays might be accounted for in this generic framework, consistently with any other constraint.
\end{abstract}

\section{Introduction}

The basic success of the Cabibbo Kobayashi Maskawa (CKM) picture of flavour and CP violations, as emerged in the last decade of experimental progress, can be summarized  as follows, as  often done in the literature. If one describes possible deviations from the CKM picture by a phenomenological effective Lagrangian of the form
\begin{equation}
\Delta \mathcal{L} = \sum_i \frac{1}{\Lambda_i^2} \mathcal{O}_i ~+\text{h.c.},
\label{bounds}
\end{equation}
where  $\mathcal{O}_i$ are gauge invariant operators of dimension 6 with generic flavour structure, in several cases the lower bounds on $\Lambda_i$ are above thousands of TeVs.

The interpretation of this fact is however far from  straightforward, since the theory of flavour, whatever it may be, will not be generic. It is for example conceivable that a flavour symmetry be operative to keep under control the coefficients in front of every operator in (\ref{bounds}). Suppose, for good reasons, that a new physics scale $\Lambda_\text{NP}$ exists somehow related to ElectroWeak Symmetry Breaking (EWSB). An ideal situation would be one such that the effective Lagrangian
\begin{equation}
\Delta \mathcal{L} = \sum_i \frac{c_i \xi_i}{\Lambda_\text{NP}^2} \mathcal{O}_i ~+\text{h.c.}
\label{ideal}
\end{equation}
is compatible with current data, where $\xi_i$ are small parameters controlled by a suitable flavour symmetry and otherwise $c_i$ are $O(1)$ coefficients. With $\Lambda_\text{NP}$  sufficiently close to the Fermi scale, this might leave room for new observable effects. Such effects would indeed be very welcome in most extensions of the SM in the EWSB sector and, if observed,  might help to shed light on a possible theory of flavour.

In previous papers \cite{Barbieri:2011ci,Barbieri:2011fc,Barbieri:2012uh} we have invoked a flavour $U(2)^3$ as a possibly relevant symmetry in this context. Such symmetry is exhibited by the quark sector of the SM, when one neglects the masses of the first two generations and their mixings with the third generation quarks. If some small breaking parameters appearing in the quark Yukawa couplings are assumed to have definite transformation properties under $U(2)^3$ and they control any other possible breaking in a putative extension of the SM, this might indeed realize the picture described above. Concrete examples are offered by supersymmetric extensions of the SM or by strongly interacting models of EWSB, e.g. with the Higgs boson as a pseudo-Goldstone particle.

In this work we want to complete the analysis of this picture by considering all the possible breaking terms of $U(2)^3$ entering the quark mass terms
\begin{equation}
\lambda_t (\qLbar \V)t_R, \quad \lambda_t \qLbar \Delta Y_u \uR,
\quad \lambda_t \bar{q}_{3L} (\Vu^\dagger \uR),
\label{Yuk_u}
\end{equation}
\begin{equation}
\lambda_b(\qLbar \V)b_R, \quad \lambda_b \qLbar \Delta Y_d \dR,
\quad \lambda_b \bar{q}_{3L} (\Vd^\dagger \dR),
\label{Yuk_d}
\end{equation}
where $\qL, \uR, \dR$  stand for doublets under $U(2)_Q, U(2)_u, U(2)_d$ respectively\footnote{In (\ref{Yuk_d}) we have factored out as  a common factor the bottom  Yukawa coupling $\lambda_b$, which in principle requires an explanation since $\lambda_b$ is relatively small. One possibility is to consider, in the massless $b$-quark limit, the symmetry 
$U(2)_Q\times U(2)_u\times SU(2)_d\times U(1)_d$, where $U(1)_d$ is the common U(1) factor acting in the same way on all the right-handed $d$-type quarks, broken by the small parameter  $\lambda_b$.}. A key assumption is that the {\it spurions} appearing in  (\ref{Yuk_u}) and (\ref{Yuk_d}) transform under $U(2)^3$ as 
\begin{equation}
\Delta Y_u = (2, 2, 1),\quad \Delta Y_d = (2, 1, 2), \quad \V = (2, 1, 1), \quad \Vu = (1, 2, 1), \quad \Vd = (1, 1, 2),
\label{spurions}
\end{equation}
such that every term in  (\ref{Yuk_u}) and (\ref{Yuk_d}) is formally invariant. 
The first two terms in (\ref{Yuk_u}) and (\ref{Yuk_d}) are minimally needed for a realistic description of quark masses and mixings. We call this Minimal $U(2)^3$.  Taking an Effective Field Theory (EFT) point of view, we have considered the current bounds on Minimal $U(2)^3$ and the corresponding new physical effects that might emerge in current flavour experiments\cite{Barbieri:2012uh}. In this paper we are interested in setting the bounds on the two remaining terms in (\ref{Yuk_u}) and (\ref{Yuk_d}) and in investigating possible new physics that these extra terms might generate. This we call Generic $U(2)^3$.

\section{Physical parameters and the CKM matrix}\label{sec:spurions}

By $U(2)^3$ transformations it is possible and useful to restrict and define the physical parameters appearing in (\ref{spurions}).  In Minimal $U(2)^3$ we choose:
\begin{equation}
\V = \begin{pmatrix}0\\ \epsilon_L\end{pmatrix},\qquad \Delta Y_u = L_{12}^u\,\Delta Y_u^{\rm diag},\qquad
 \Delta Y_d = \Phi_L L_{12}^d\,\Delta Y_d^{\rm diag},
 \end{equation}
 where $\epsilon_L$ is a real parameter, $L_{12}^{u,d}$ are rotation matrices in the space of the first two generations with angles $\theta_L^{u,d}$ and $\Phi_L = {\rm diag}\big(e^{i\phi},1\big)$, i.e. four parameters in total.
 
 Similarly in Generic $U(2)^3$ we set:
 \begin{equation}\label{genericV}
\V = \begin{pmatrix}0\\ \epsilon_L\end{pmatrix},\qquad \Vu = \begin{pmatrix}0\\ \epsilon^u_R\end{pmatrix},\qquad \Vd = \begin{pmatrix}0\\ \epsilon_R^d\end{pmatrix},
\end{equation}
\begin{align}\label{genericY}
\Delta Y_u &= L_{12}^u\,\Delta Y_u^{\rm diag}\,\Phi_R^u R_{12}^u, & \Delta Y_d &= \Phi_L L_{12}^d\,\Delta Y_d^{\rm diag}\,\Phi_R^d R_{12}^d,\\
\Phi_L &= {\rm diag}\big(e^{i\phi},1\big), & \Phi_R^{u, d} &= {\rm diag}\big(e^{i\phi_1^{u,d}}, e^{i\phi_2^{u,d}}\big),
\end{align}
 which adds to the four parameters of Minimal $U(2)^3$ four real parameters, $\epsilon_R^{u,d}, \theta_R^{u,d}$ and four phases, $\phi_{1,2}^{u,d}$. For later convenience we define $s^{u,d}_L=\sin{\theta_L^{u,d}}$ and $s^{u,d}_R=\sin{\theta_R^{u,d}}$.
 
 The next step consists in writing down the mass terms for the up and down-type quarks, invariant under  $U(2)^3$, and in diagonalizing them\footnote{See Appendix \ref{app:bilinears} for a detailed analysis of their digonalization, together with all the possible quark bilinears appearing in effective operators.}, which can be done perturbatively by taking into account the smallness of $\epsilon_L, \epsilon_R^{u,d}$ and  $\Delta Y_{u, d}^{\rm diag}$. As a consequence, to a sufficient approximation the unitary transformations that bring these mass matrices to diagonal form are influenced on the left side only by the four parameters of  Minimal $U(2)^3$, $\epsilon_L, \theta_L^{u,d}, \phi$, whereas those on the right side depend on the extra parameters of Generic $U(2)^3$, $\epsilon_R^{u,d}, \theta_R^{u,d}, \phi_{1,2}^{u,d}$. In turn this leads to a unique form of the standard CKM matrix
 \begin{equation}\label{CKM}
V_{CKM} = \begin{pmatrix}
c^u_L c^d_L & \lambda & s^u_L s\,e^{-i\delta}\\
-\lambda & c^u_L c^d_L & c^u_L s\\
-s^d_L s\,e^{i(\delta - \phi)} & -c^d_L s & 1
\end{pmatrix},
\end{equation}
where $s\sim O(\epsilon_L)$, $c^{u,d}_L=\cos{\theta_L^{u,d}}$ and
$s^u_L c^d_L - s^d_L c^u_L e^{i\phi} = \lambda e^{i\delta}$.
Using this parametrization of the CKM matrix, a direct fit of the tree-level flavour observables, presumably not influenced by new physics,  results in \cite{Barbieri:2011fc}
\begin{align}
s^u_L &= 0.086\pm0.003
\,,&
s^d_L &= -0.22\pm0.01
\,,\\
s &= 0.0411 \pm 0.0005
\,,&
\phi &= (-97\pm 9)^\circ
\,.
\end{align}
At this stage, the extra ``right-handed'' parameters present in Generic $U(2)^3$ are unconstrained.

\section{Effective Field Theory analysis}

As outlined in the Introduction, we are interested in considering from an EFT point of view the leading flavour-violating  operators that are consistent with the $U(2)^3$ symmetry, only broken by the {\it spurions} in (\ref{Yuk_u}) and (\ref{Yuk_d}). Their general form can be summarized in
\begin{equation}
\Delta \mathcal{L} = \Delta \mathcal{L}^{4f}_{L}  + \Delta \mathcal{L}_\text{mag}  + \Delta \mathcal{L}^{4f}_{R}  + \Delta \mathcal{L}^{4f}_{LR}\, ,
\label{eq:genL}
\end{equation}
 where $\Delta \mathcal{L}^{4f}_{L, R, LR} $ are the sets of four-fermion operators with flavour violation respectively in the left-handed sector, in the right-handed sector and in both, whereas $\Delta \mathcal{L}_\text{mag}$ contains the chirality-breaking dipole operators. Notice that sizeable contributions to $\Delta \mathcal{L}^{4f}_{R}$ and $\Delta \mathcal{L}^{4f}_{LR}$ are absent in Minimal $U(2)^3$.
 
 In what follows, we will write each single term in \eqref{eq:genL} as
\begin{equation}
\Delta \mathcal{L}_{} = \frac{1}{\Lambda^2} \sum_i C_i \mathcal{O}_i ~+\text{h.c.}
,
\end{equation}
where the operators $\mathcal{O}_i$ relevant for the process under examination will be specified case by case.

\subsection{Flavour and CP violation in Minimal $U(2)^3$}\label{sec:minimal}
 
 Already Minimal $U(2)^3$ gives full rise to $\Delta \mathcal{L}^{4f}_{L}$ as well as to the  $b_R\rightarrow s_L (d_L)$ transitions in $\Delta \mathcal{L}_\text{mag}$. The bounds on the coefficients of these operators arise from $\Delta F=2$ transitions, $\epsilon_K$, $B_d^0$-$\bar{B}_d^0$ mixing, $B_s^0$-$\bar{B}_s^0$ mixing, and from $B$ decays, mostly $b\rightarrow s \gamma$, $b\rightarrow s \ell\bar{\ell}$, $b\to s\nu\bar{\nu}$. The  analysis in \cite{Barbieri:2012uh} shows that the operators in $\Delta \mathcal{L}^{4f}_{L}$ and $\Delta \mathcal{L}_\text{mag}$, controlled by Minimal $U(2)^3$  breaking, are broadly consistent with an overall scale at 3 TeV and coefficients $c_i$ in the range $0.2$ to $1$, depending on their phases.
 
 \subsubsection{Analysis of $\epsilon'/\epsilon$} 
An observable that has not been included in \cite{Barbieri:2012uh} and is actually relevant in other contexts as well, like in Minimal Flavour Violation (MFV) \cite{D'Ambrosio:2002ex}, is direct CP violation in $K$ decays, as summarized in the parameter $\epsilon^\prime$.
 Either in $U(2)^3$ or in MFV, a contribution to $\epsilon^\prime$ arises from the operators
 \begin{equation}\label{operatorsepsilonprime}
 \Delta \mathcal{L}^{4f, \Delta S = 1}_{LR} = \frac{1}{\Lambda^2} \xi_{ds} (c^d_5 \mathcal{O}^d_5 + c^u_5 \mathcal{O}^u_5 + c^d_6 \mathcal{O}^d_6 + c^u_6 \mathcal{O}^u_6) + \text{h.c.}, ~\xi_{ds} = V_{td}V_{ts}^*\, ,
 \end{equation}
 where
 \begin{align}
 \mathcal{O}_5^q = ( \bar{d}_L \gamma_\mu s_L ) (\bar{q}_R \gamma_\mu q_R), \qquad
 \mathcal{O}_6^q = ( \bar{d}_L^\alpha \gamma_\mu s_L^\beta) ( \bar{q}_R^\beta \gamma_\mu q_R^\alpha),\qquad q=u,d.
\end{align}

 The dominant contribution to the $\epsilon'$ parameter reads
\begin{equation}
\left|\frac{\epsilon'}{\epsilon}\right| \simeq \frac{\left| \text{Im}A_2 \right|}{\sqrt{2} \, |\epsilon| \,\text{Re}A_0} \,,
\end{equation}
where $A_i=A(K \to (\pi \pi)_{I=i})$. 
Using $\langle (\pi \pi)_{I=2} | \mathcal{O}_i^u + \mathcal{O}_i^d | K \rangle \simeq 0$ from isospin conservation and neglecting contributions from other operators, which are subleading,  we obtain
\begin{align}
 \text{Im}A_2 = \frac{1}{\Lambda^2}\left[ \left( C_5^d-C_5^u\right) \langle (\pi \pi)_{I=2} | \mathcal{O}_5^d| K \rangle + \left( C_6^d-C_6^u\right) \langle (\pi \pi)_{I=2} | \mathcal{O}_6^d| K \rangle \right].
\end{align}
From \cite{Bosch:1999wr} we have at the scale $\mu = m_c$
\begin{align}
\langle (\pi \pi)_{I=2} | \mathcal{O}_5^d | K \rangle & \simeq -\frac{1}{6 \sqrt{3}}\big( m_K^2 \rho^2 - m_K^2 + m_\pi^2\big) f_\pi  B_7^{(3/2)}(m_c), \nonumber \\
\langle (\pi \pi)_{I=2} | \mathcal{O}_6^d | K \rangle & \simeq -\frac{1}{2 \sqrt{3}} \big( m_K^2 \rho^2 - \frac{1}{6} ( m_K^2 - m_\pi^2) \big)  f_\pi B_8^{(3/2)}(m_c) ,
\end{align}
where $\rho = m_K/m_s$. In the following we set $B_7^{(3/2)}(m_c) =  B_8^{(3/2)}(m_c) =1$.

The coefficients $C^{(3/2)}_i = C_i^d-C_i^u$ at the low scale $\mu$ read in terms of those at the high scale $\Lambda$\cite{Buchalla:1995vs}
\begin{align}
C^{(3/2)}_5 (m_c) &= \eta_5 C^{(3/2)}_5 (\Lambda), \nonumber \\
C^{(3/2)}_6 (m_c) &= \eta_6 C^{(3/2)}_6(\Lambda) + \frac{1}{3}(\eta_6 - \eta_5) C^{(3/2)}_5(\Lambda),
\end{align}
where
\begin{align}
\eta_5 &= \left(\frac{\alpha_s(\Lambda)}{\alpha_s(m_t)}\right)^{\frac{3}{21}} \left(\frac{\alpha_s(m_t)}{\alpha_s(m_b)}\right)^{\frac{3}{23}}\left(\frac{\alpha_s(m_b)}{\alpha_s(m_c)}\right)^{\frac{3}{25}}  \simeq 0.82\,,\notag \\
\eta_6 &= \left(\frac{\alpha_s(\Lambda)}{\alpha_s(m_t)}\right)^{-\frac{24}{21}} \left(\frac{\alpha_s(m_t)}{\alpha_s(m_b)}\right)^{-\frac{24}{23}} \left(\frac{\alpha_s(m_b)}{\alpha_s(m_c)}\right)^{-\frac{24}{25}} \simeq 4.83 \,.
\end{align}

Requiring the extra contribution from $\Delta \mathcal{L}^{4f, \Delta S = 1}_L$ to respect $|\epsilon^\prime/\epsilon | < |\epsilon^\prime/\epsilon |_\text{exp} \simeq 1.7\times 10^{-3}$, we obtain 
\begin{align}\label{boundepsilonprime}
c_5^{u,d} &\lesssim 0.4 \left(\frac{\Lambda}{3~\text{TeV}}\right)^2, & c_6^{u,d} &\lesssim 0.13 \left(\frac{\Lambda}{3~\text{TeV}}\right)^2.
\end{align}
Taking into account the uncertainties in the estimate of the SM contribution to $\epsilon^\prime/\epsilon$, which could cancel against a new physics contribution, as well as the uncertainties in the $B_i$ parameters, this bound might perhaps be relaxed by a factor of a few\footnote{Note that in Supersymmetry the heaviness of the first generation squark circulating in the box loop suppresses the coefficients $c_5^{u,d}$ and $c_6^{u,d}$.}.

\subsubsection{Electric dipole moment of the neutron}
Already in \cite{Barbieri:2012uh} we remarked that the presence of phases in flavour-diagonal chirality breaking operators has to be consistent with the limits coming from the neutron electric dipole moment (EDM). Here we make the statement more precise, deriving a quantitative bound for the Minimal breaking case, also for comparison with  the General breaking case,  analyzed in the next section.

The relevant contraints come from the CP violating contributions to the operators
\begin{align}\label{EDM}
\Delta\mathcal{L}_\text{mag}^{\Delta F=0} &= \frac{1}{\Lambda^2} \left[
\tilde c_u^g e^{i\tilde \phi_u^g} m_u(\bar u_L\sigma_{\mu\nu}T^a u_R)
+ \tilde c_u^\gamma e^{i\tilde\phi_u^\gamma} m_d(\bar d_L\sigma_{\mu\nu}T^a d_R)
\right] g_s G^{\mu\nu}_a \nonumber \\
 &+
 \frac{1}{\Lambda^2} \left[
\tilde c_d^g e^{i\tilde\phi_d^g} m_u(\bar u_L\sigma_{\mu\nu} u_R)
+ \tilde c_d^\gamma e^{i\tilde\phi_d^\gamma} m_d(\bar d_L\sigma_{\mu\nu} d_R)
\right] eF^{\mu\nu} +\text{h.c.}\, ,
\end{align}
where we have made all the phases explicit. 
In terms of the coefficients of \eqref{EDM}, the up and down quark electric dipole moments (EDMs) and chromoelectric dipole moments (CEDMs), defined as in \cite{Pospelov:2000bw}, are
\begin{align}
d_{q} = 2 e \, \frac{m_q}{\Lambda^2}\,\tilde c_{q}^{\gamma}\sin(\tilde\phi_i^{\gamma}),\qquad \tilde d_q = 2 \, \frac{m_q}{\Lambda^2}\, \tilde c_q^g\sin(\tilde \phi_q^g),\qquad q=u,d.
\end{align}
The contribution to the neutron EDM reads \cite{Pospelov:2000bw}
\begin{equation} \label{neutron}
d_n = (1 \pm 0.5) \left( 1.4( d_d - \tfrac{1}{4} d_u) + 1.1 e ( \tilde{d}_d + \tfrac{1}{2} \tilde{d}_u)  \right) \, ,
\end{equation}
where all the coefficients are defined at a hadronic scale of 1~GeV.

Taking into account the RG evolution between 3~TeV and the hadronic scale, the 90\% C.L. experimental bound
$|d_n| < 2.9 \times 10^{-26} ~e\,\text{cm}$ \cite{Baker:2006ts}
implies for the parameters at the high scale
\begin{align}
\tilde c_{u}^{\gamma}\sin(\tilde\phi_u^{\gamma}) &\lesssim 1.9 \times 10^{-2}\left(\frac{\Lambda}{3~\text{TeV}}\right)^2, & \tilde c_{d}^{\gamma}\sin(\tilde\phi_d^{\gamma}) \lesssim 2.4 \times 10^{-3}\left(\frac{\Lambda}{3~\text{TeV}}\right)^2,\\
\tilde{c}_{u}^g\sin(\tilde\phi_u^{g}) &\lesssim 7.1 \times 10^{-3}\left(\frac{\Lambda}{3~\text{TeV}}\right)^2, & \tilde{c}_{d}^g\sin(\tilde\phi_d^{g}) \lesssim 1.8 \times 10^{-3}\left(\frac{\Lambda}{3~\text{TeV}}\right)^2.
\end{align}
Note that the bounds are automatically satisfied if one does not allow for phases outside the spurions. Otherwise, even with generic phases $\tilde \phi_{u,d}^{\gamma,g}$, the smallness of the coefficients can be explained taking the new physics scale related to the first generation decoupled from the scale $\Lambda$ of EWSB, as allowed by the $U(2)^3$ symmetry.
This can be realized in concrete models such as supersymmetry, where the operators in \eqref{EDM} come from Feynman diagrams involving the exchange of heavy first generation partners.

 \subsection{Flavour and CP violation in Generic  $U(2)^3$ }
 
Generic $U(2)^3$, introducing physical rotations in the right handed sector as well, gives rise to extra flavour and CP violating contributions in \eqref{eq:genL}. The most significant of them are contained in  $\Delta \mathcal{L}_\text{mag}$ and in $\Delta \mathcal{L}^{4f}_{LR}$. In the following, we first discuss the relevant new effects with respect to Minimal~$U(2)^3$, which show up in $\Delta C = 1$, $\Delta S = 1$ and $\Delta S = 2$ observables, as well as in flavour conserving electric dipole moments. We then see how in $B$ and $t$ decays and in $D$-$\bar{D}$ mixing the new effects are at most analogous in magnitude to those of the Minimal breaking case.

\subsubsection{$\Delta C=1$: $D$ decays}

CP asymmetries in $D$ decays receive contributions from chromo-magnetic dipole operators with both chiralities,
\begin{equation}
\mathcal{L}^{\Delta C=1}_\text{mag} = \frac{1}{\Lambda^2} c_D^g e^{i \phi_D^g}
\zeta_{uc} \left[
e^{-i\phi_2^u} \frac{\epsilon^u_R}{\epsilon_L}\, \mathcal{O}_8
+
e^{i\phi_1^u} \frac{s^u_R}{s^u_L} \frac{\epsilon_R^u}{\epsilon_L}\, \mathcal{O}_8'
\right] + \text{h.c.}\\
\end{equation}
where
\begin{equation}
\mathcal{O}_8 = m_t(\bar u_L\sigma_{\mu\nu}T^a c_R)g_s G^{\mu\nu}_a,\qquad \mathcal{O}_8' = m_t(\bar u_R\sigma_{\mu\nu}T^a c_L)g_s G^{\mu\nu}_a,
\end{equation}
and with $\phi_D^{g}$ we account for the possibility of CP violating phases outside the spurions (see Appendix \ref{app:bilinears} for details). 
Most notably, the recently observed CP asymmetry difference between $D\to KK$ and $D\to \pi\pi$ decays could be due to new physics contributions to the chromo-magnetic operators. Following \cite{Isidori:2011qw,Giudice:2012qq} we write at the scale $\mu = m_c$
\begin{equation}
\Delta A_\text{CP} \simeq - \frac{2}{\lambda} \Big[\text{Im}(V_{cb}^* V_{ub}) \text{Im}\left(\Delta R^{\text{SM}} \right) + \frac{1}{\Lambda^2}
\Big(
\text{Im}\big( C_8\big) \text{Im}\big( \Delta R^{\text{NP}} \big) +
\text{Im} ( C_8' ) \text{Im}\big( {\Delta R'}^{\text{NP}} \big) 
\Big) \Big]
,
\label{DeltaAcp}
\end{equation}
where the Cabibbo angle $\lambda$ is defined in \eqref{CKM}, $\Delta R^{(\prime)\text{SM},\text{NP}} = R_K^{(\prime)\text{SM},\text{NP}} + R_{\pi}^{(\prime)\text{SM}, \text{NP}}$, $R_{K,\pi}^{\text{SM}}$ are the ratios between the subleading and the dominant SM hadronic matrix elements, and
\begin{align}
\label{Dhadr}
{R_{K}^{(\prime)\text{NP}}} &\simeq V_{cs}^* V_{us} \frac{\langle K^+ K^- | \mathcal{O}_8^{(\prime)}| D\rangle}{\langle K^+ K^- | \mathcal{L}^{\text{SM}}_{eff}| D \rangle} \sim 0.1 \times \frac{4 \pi^2 m_t}{m_c} \frac{\sqrt{2}}{G_F},\\
R_{\pi}^{(\prime)\text{NP}} &\simeq V_{cd}^* V_{ud} \frac{\langle \pi^+ \pi^- | \mathcal{O}_8^{(\prime)}| D\rangle}{\langle \pi^+ \pi^- | \mathcal{L}^{\text{SM}}_{eff}| D \rangle} \simeq R_K^{(\prime)\text{NP}}.
\end{align}
In our estimates we will assume maximal strong phases, which imply $|\text{Im}\, \Delta R^{(\prime)\text{NP}}| \simeq 2 R_{K}^{\text(\prime){NP}}$. The SM contribution can be naively estimated to be $\Delta R^{\text{SM}} \sim \alpha_s(m_c)/\pi \sim 0.1$, but larger values from long distance contributions could arise, 
making a possible explanation within the SM  still an open issue (see e.g. \cite{Brod:2012ud,Isidori:2012yx} for recent works on this).

Requiring the new physics contribution to $\Delta A_\text{CP}$ to be less than the experimental central value of the combination between LHCb \cite{Aaij:2011in} and CDF \cite{CDF-Note-10784} , $\Delta A_\text{CP}^\text{exp}=(-0.67\pm0.16)\%$, 
implies
\begin{align}
c_D^{g}\, \frac{\epsilon^u_R}{\epsilon_L}\, \frac{\sin\left(\delta -\phi_2^u + \phi_D^{g}\right)}{\sin \delta}
&\lesssim 0.35
\left( \frac{\Lambda}{3\, {\rm TeV}} \right)^2
,
&
c_D^{g}\, \frac{s^u_R}{s^u_L} \frac{\epsilon^u_R}{\epsilon_L}\, \frac{\sin(\delta + \phi_1^u - \phi_D^{g})}{\sin \delta}
&\lesssim 0.35
\left( \frac{\Lambda}{3\, {\rm TeV}} \right)^2
.
\label{eq:DC1bound}
\end{align}
The bound can be saturated without violating indirect constraints on these operators arising from $\epsilon'$ or $D$-$\bar D$ mixing due to weak operator mixing \cite{Isidori:2011qw}. We stress that the bounds in \eqref{eq:DC1bound} carry an order one uncertainty coming from the normalized matrix elements $R_{\pi,K}$.

\subsubsection{$\Delta F=0$: neutron EDM}

In the flavour conserving case, important constraints arise from the up and down quark electric dipole moments (EDMs) and chromo-electric dipole moments (CEDMs). In addition to \eqref{EDM} there are new contributions coming from the CP violating part of the operators
\begin{align}\label{EDMright}
 \mathcal{L}^{\Delta F=0}_\text{mag} &= \frac{m_t}{\Lambda^2} \xi_{uu}\,e^{-i \phi_1^u} \frac{s^u_R}{s^u_L} \frac{\epsilon^u_R}{\epsilon_L}\left[
c_u^g e^{i\phi_u^g} (\bar u_L\sigma_{\mu\nu}T^a u_R)g_sG^{\mu\nu}_a
+ c_u^\gamma e^{i\phi_u^\gamma} (\bar u_L\sigma_{\mu\nu}u_R)eF^{\mu\nu}
\right] \nonumber \\
 &+
 \frac{m_b}{\Lambda^2} \xi_{dd} \,e^{-i \phi_1^d} \frac{s^d_R}{s^d_L} \frac{\epsilon^d_R}{\epsilon_L}\left[
c_d^g e^{i\phi_d^g} (\bar d_L\sigma_{\mu\nu}T^a d_R)g_sG^{\mu\nu}_a
+ c_d^\gamma e^{i\phi_d^\gamma} (\bar d_L\sigma_{\mu\nu}d_R)eF^{\mu\nu}
\right] + \text{h.c.}\, ,
\end{align}
where we remind that $\phi_1^{u,d}$ are non zero even if there are no CP phases outside the spurions. The new contributions to the quark (C)EDMs are
\begin{equation}
d_u = 2e\frac{m_t}{\Lambda^2}\xi_{uu}\frac{s^u_R}{s^u_L}\frac{\epsilon^u_R}{\epsilon_L}c_u^{\gamma}\sin(\phi_u^{\gamma} - \phi_1^u),\qquad \tilde d_u = 2\frac{m_t}{\Lambda^2}\xi_{uu} \frac{s^u_R}{s^u_L}\frac{\epsilon^u_R}{\epsilon_L}c_u^{g}\sin(\phi_u^{g} - \phi_1^u), \qquad (u\leftrightarrow d).
\end{equation}
From \eqref{neutron}, considering again the running of the Wilson coefficients from 3 TeV down to the hadronic scale of 1 GeV, the experimental bound on the neutron EDM implies for the parameters at the high scale
\begin{align}
c_u^{\gamma}\,|\sin (\phi_u^{\gamma}-\phi_1^u )| \frac{s^u_R}{s^u_L}\,\frac{\epsilon^u_R}{\epsilon_L}\lesssim 1.2 \times 10^{-2}
\left( \frac{\Lambda}{3\, {\rm TeV}} \right)^2
,
\nonumber\\
c_d^{\gamma}\,|\sin (\phi_d^{\gamma}-\phi_1^d )|  \frac{s^d_R}{s^d_L}\,\frac{\epsilon^d_R}{\epsilon_L}\lesssim 3.2 \times 10^{-2}
\left( \frac{\Lambda}{3\, {\rm TeV}} \right)^2
,
\nonumber\\
c_u^{g}\,|\sin (\phi_u^{g}-\phi_1^u )|  \frac{s^u_R}{s^u_L}\,\frac{\epsilon^u_R}{\epsilon_L}\lesssim 4.4 \times 10^{-3}
\left( \frac{\Lambda}{3\, {\rm TeV}} \right)^2
,
\nonumber\\
c_d^{g}\,|\sin (\phi_d^{g}-\phi_1^d )|  \frac{s^d_R}{s^d_L}\,\frac{\epsilon^d_R}{\epsilon_L}\lesssim 2.5 \times 10^{-2}
\left( \frac{\Lambda}{3\, {\rm TeV}} \right)^2
.
\label{eq:DF0bound}
\end{align}

Notice that since the operators of \eqref{EDMright} are generated through the right-handed mixings with the third generation, the coefficients $c_{u,d}^{\gamma, g}$ can no longer be suppressed by the large mass of first-generation-partners as in the Minimal case, and the bounds above will constrain $s^u_R\epsilon^u_R$ and $s^d_R\epsilon^d_R$.

\subsubsection{$\Delta S=1$: $\epsilon'/\epsilon$}

The $s\to d$ chromomagnetic dipole
\begin{align}
 \Delta\mathcal{L}^{\Delta S = 1}_\text{mag} &= 
 \frac{m_t}{\Lambda^2} {c}_K^g e^{i(\phi_K^g-\phi_2^d)}\lambda_b \xi_{ds} \frac{\epsilon^d_R}{\epsilon_L} \left( \bar{d}_L \sigma_{\mu\nu} T^a s_R \right) g_s G_{\mu\nu}^a
\end{align}
contributes to $\epsilon'$. Following the analysis in \cite{Mertens:2011ts}, one obtains the bound
\begin{equation}
{c}_K^g \frac{\sin (\beta +\phi_K^g - \phi_2^d)}{\sin \beta}\,\frac{\epsilon^d_R}{\epsilon_L}
\lesssim 0.7 \left(\frac{\Lambda}{3 \,{\rm TeV}} \right)^2 \,.
\label{eq:DS1bound}
\end{equation}
Furthermore, in addition to the LR four fermion operators in \eqref{operatorsepsilonprime}, there is a contribution to $\epsilon'$ also from the operators with exchanged chiralities
\begin{equation}
\Delta \mathcal{L}^{4f, \Delta S = 1}_{LR} = \frac{1}{\Lambda^2} \xi_{ds} \frac{s^d_R}{s^d_L} \left(\frac{\epsilon^d_R}{\epsilon_L}\right)^2 e^{i(\phi_1^d-\phi_2^d)}(c^{\prime d}_5 \mathcal{O}^{\prime d}_5 + c^{\prime u}_5 \mathcal{O}^{\prime u}_5 + c^{\prime d}_6 \mathcal{O}^{\prime d}_6 + c^{\prime u}_6 \mathcal{O}^{\prime u}_6) + \text{h.c.}\, ,
\end{equation}
where
\begin{align}
 \mathcal{O}_5^{\prime q} = ( \bar{d}_R \gamma_\mu s_R ) (\bar{q}_L \gamma_\mu q_L), \qquad
 \mathcal{O}_6^{\prime q} = ( \bar{d}_R^\alpha \gamma_\mu s_R^\beta) ( \bar{q}_L^\beta \gamma_\mu q_L^\alpha),\qquad q=u,d.
\end{align}
From \eqref{boundepsilonprime} one gets
\begin{align}
c_5^{\prime u,d} \frac{\sin (\beta +\phi_1^d - \phi_2^d)}{\sin \beta}\, \frac{s^d_R}{s^d_L} \left(\frac{\epsilon^d_R}{\epsilon_L}\right)^2
&\lesssim 0.4\left(\frac{\Lambda}{3~\text{TeV}}\right)^2,\\
c_6^{\prime u,d} \frac{\sin (\beta +\phi_1^d - \phi_2^d)}{\sin \beta}\, \frac{s^d_R}{s^d_L} \left(\frac{\epsilon^d_R}{\epsilon_L}\right)^2
&\lesssim 0.13\left(\frac{\Lambda}{3~\text{TeV}}\right)^2,
\end{align}
which is not particularly relevant since a stronger bound on the combination $(s^d_R/s^d_L)(\epsilon^d_R/\epsilon_L)^2$ comes from $\epsilon_K$.

\subsubsection{$\Delta S=2$: $\epsilon_K$}
   
Finally, the only relevant new effect contained in $\Delta \mathcal{L}^{4f}_{LR}$ arises from $\Delta S=2$ operators contributing to $\epsilon_K$, which are enhanced by a chiral factor and by renormalization group effects. The relevant operators are
\begin{equation}
\Delta\mathcal{L}^{\Delta S=2}_{LR} = \frac{1}{\Lambda^2}\frac{s^d_R}{s^d_L} \left(\frac{\epsilon^d_R}{\epsilon_L}\right)^2\xi_{ds}^2 e^{i(\phi_1^d-\phi_2^d)}\left[ c_K^{SLR}\lambda_b^2  \left(\bar{d}_L s_R\right) \left(\bar{d}_R s_L \right) + c_K^{VLR} \left(\bar{d}_L \gamma_\mu s_L \right)\left(\bar{d}_R \gamma_\mu s_R \right)\right]
.
\end{equation}
Using bounds from \cite{Isidori:2010kg}, one gets
\begin{equation}
 c_K^{VLR} \frac{\sin (2 \beta +\phi_1^d - \phi_2^d)}{\sin 2 \beta}\, \frac{s^d_R}{s^d_L} \left(\frac{\epsilon^d_R}{\epsilon_L}\right)^2
\lesssim 6 \times 10^{-3} \left(\frac{\Lambda}{3 \,{\rm TeV}} \right)^2\,.
\label{eq:DS2bound}
\end{equation}

\subsubsection{$D$ mixing, $B$ and top FCNCs}

In the $D$ and $B$ systems there are no enhancements of the matrix elements of the operators in $\Delta \mathcal{L}_{LR}^{4f}$ and $\Delta \mathcal{L}_{R}^{4f}$, unlike what happens for $K$ mesons. Moreover the new contributions to these operators are all suppressed by some powers of $\epsilon^{u,d}_R/\epsilon_L$ (see Appendix \ref{app:bilinears}). Therefore they are all subleading with respect to those of Minimal $U(2)^3$, once we take into account the bounds from the other observables that we have discussed. An analogous suppression holds also for the operators that contain chirality breaking bilinears involving one third generation quark, relevant for $B$ and top FCNCs. 
A four fermion operator of the form $(\bar{u}_
L c_R)(\bar{u}_R c_L)$
might in principle be relevant for $D$-$\bar{D}$ mixing. However, taking into account the bounds on $\epsilon^u_R/\epsilon_L$,  this new contribution gives effects of the same size of those already present in Minimal $U(2)^3$ and far from the current sensitivity.
Consequently the phenomenology of $B$ decays is the same for Minimal and Generic $U(2)^3$. The only difference  in the latter is that CP violating effects are generated also if we set to zero the phases outside the spurions, though suppressed by at least one power of $\epsilon^d_R/\epsilon_L$.

Concerning the up quark sector, 
given the future expected sensitivities for top FCNCs \cite{Carvalho:2007yi} and CPV in $D$-$\bar{D}$ mixing \cite{Aushev:2010bq, Merk:2011zz}, within the $U(2)^3$ framework we continue to expect no significant effects in these processes (see \cite{Barbieri:2012uh} for the size of the largest contributions). We stress that, while an observation of a flavour changing top decay at LHC would generically put the $U(2)^3$ framework into trouble, a hypothetical observation of CP violation in $D$-$\bar{D}$ mixing would call for a careful discussion of the long distance contribution.

\subsubsection{Comparison of bounds}

\begin{figure}[tb]
\centering
\includegraphics[width=0.8\textwidth]{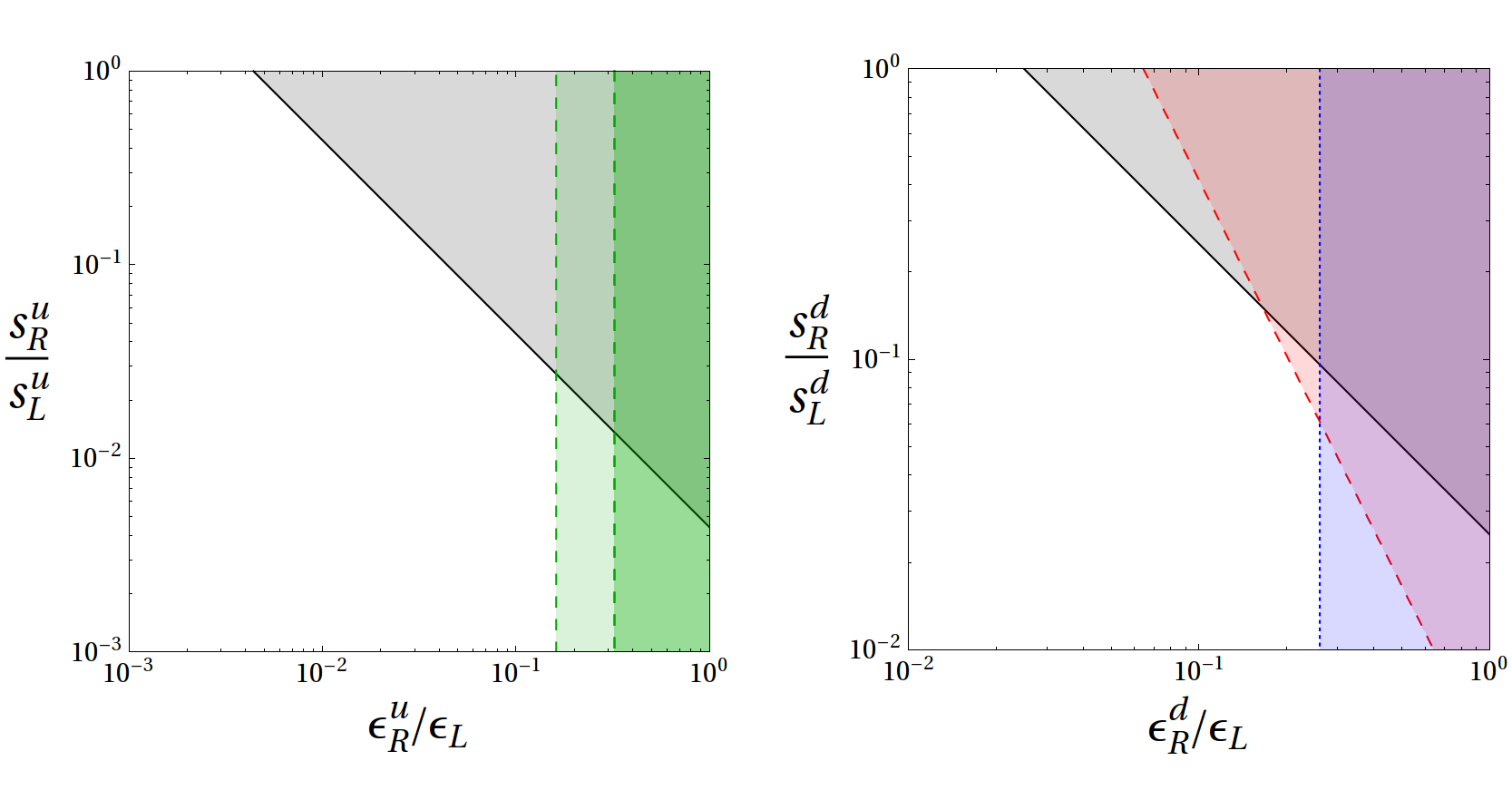}
\caption{Bounds on the free parameters of Generic $U(2)^3$ breaking, normalized to the parameters present in Minimal $U(2)^3$ breaking (determined by the CKM), with maximal phases.
The black solid line in both plots shows the bound from the neutron EDM (the shaded region is disfavoured at 90\% C.L.). In the left-hand plot, the green dashed lines correspond to a new physics contribution to $\Delta A_\text{CP}$ of 50\% and 100\% of the experimental central value. The darker shaded region is disfavoured, while in the lighter region, new physics could account for the large experimental value. In the right-hand plot, the red dashed line shows the bound from $\epsilon_K$ and the blue dotted line the one from $\epsilon'$.
}
\label{fig:bounds}
\end{figure}

The bounds in eqs. (\ref{eq:DC1bound}), (\ref{eq:DF0bound}), (\ref{eq:DS1bound}) and (\ref{eq:DS2bound}) constrain the $U(2)^3$ breaking parameters $\epsilon^{u,d}_R$ and $s^{u,d}_R$ for given values of the model-dependent parameters $c_i^{\alpha}$ and phases. Assuming all the real parameters to be unity and all the phases to be such as to maximize the corresponding bounds on the $U(2)^3$ breaking parameters, to be conservative, fig.~\ref{fig:bounds} compares the strength of the bounds from the different observables. One can make the following observations:
\begin{itemize}
\item $\Delta A_\text{CP}$ could be due to new physics compatible with $U(2)^3$ if $\epsilon^u_R\sim0.1\epsilon_L$. However, with phases that maximize all the constraints, the bound on the up-quark CEDM then requires the angle $s^u_R$ to be more than one order of magnitude smaller than the corresponding ``left-handed'' angle $s^u_L$, whose size is determined by the CKM matrix.
\item If $\epsilon^d_R\lesssim0.1\epsilon_L$, bounds from the kaon system and the down quark (C)EDM are satisfied even without a considerable alignment of the $\Delta Y_d$ spurion.
\end{itemize}
Needless to say, in concrete models the relative strength of these bounds could vary by factors of a few.

\section{Summary and conclusions}

\begin{table}[tb]
\renewcommand{\arraystretch}{1.5}
 \begin{center}
\begin{tabular}{lcccccc}
&\multicolumn{2}{c}{Chirality conserving} & \multicolumn{2}{c}{Chirality breaking}\\
\hline
& $\Delta B = 1,2$ & $\Delta S = 1,2$ & $\Delta B = 1$ & $\Delta C = 1$ 
\\\hline
$U(3)^3$ moderate $t_\beta$ & \multicolumn{2}{l}{\hspace{.61cm}\ovalbox{$\mathbbm R\qquad\qquad\;\; \mathbbm R$}} & $\mathbbm C$ & 0 
\\
Minimal $U(2)^3$, $U(3)^3$ large $t_\beta$  & $\mathbbm C$ & $\mathbbm R$ & $\mathbbm C$ & 0 
\\
Generic $U(2)^3$ & $\mathbbm C$ & $\mathbbm C$ & $\mathbbm C$ & $\mathbbm C$ 
\\\hline
 \end{tabular}
 \end{center}
\caption{Expected new physics effects in $U(3)^3$ and both Minimal and Generic $U(2)^3$, for chirality conserving and chirality breaking $\Delta F=1,2$ FCNC operators in the $B$, $K$, $D$ systems. $\mathbbm R$ denotes possible effects, but aligned in phase with the SM, $\mathbbm C$ denotes possible effects with a new phase, and 0 means no or negligible effects. In $U(3)^3$ with moderate $\tan\beta$ an additional feature is that the effects in $b\to q$ ($q = d, s$) and $s\to d$ transitions are perfectly correlated.
}
\label{tab:u2u3}
\end{table} 
 
A suitably broken $U(2)^3$ flavour symmetry may allow for deviations from the CKM picture of flavour and CP violations related to new physics at the ElectroWeak scale and waiting to be discovered. 
We have defined a Minimal and a Generic $U(2)^3$ case, depending on the fact that one takes a minimal set of breaking {\it spurions} or one allows all the possible terms contributing to the quark mass terms. 
Using an EFT approach to Minimal $U(2)^3$ it is possible to write down an effective Lagrangian
 \begin{equation}
\Delta \mathcal{L} = \sum_i \frac{c_i \xi_i}{(4\pi v)^2} \mathcal{O}_i ~+\text{h.c.}
\label{ideal2}
\end{equation}
where the $\xi_i$ are suitable combinations of the standard CKM matrix elements and $|c_i| = 0.2$ to $1$ consistently with current experimental constraints. This remains true even after the inclusion of the constraint coming from direct CP violation in K decays (the $\epsilon^\prime$ parameter) which had escaped  attention so far, to the best of our knowledge, in the EFT context. If these considerations are of any guidance,  the main observables that deserve attention, in view of conceivable experimental progress, are CP violation in the mixing of the $B_s$ system, rates and/or asymmetries in $B$ decays, like $b\rightarrow s(d)\gamma$, $b\rightarrow s(d) \ell\bar{\ell}$, $b\to s(d)\nu\bar{\nu}$  and in $K\rightarrow \pi \nu\bar{\nu}$ decays, either charged or neutral. 
 
Generic $U(2)^3$ introduces new parameters which do not have a correspondence with the ones of standard CKM. 
As such, even insisting on an effective scale at 3 TeV, one cannot predict the size of the extra effects introduced in  Generic $U(2)^3$. We have seen, however,  where the main constraints on the new parameters come from: in the up sector from  CP asymmetries in $D$ decays and from the neutron EDM and, in the down sector, also from the neutron EDM and from CP violation in the Kaon system. Always with an effective scale at 3 TeV and barring cancellations among  phases, the size of the new breaking terms included in Generic $U(2)^3$ have to be smaller than the corresponding ones in Minimal $U(2)^3$. It is for example nevertheless possible to attribute the recently measured CP violation in $D~\rightarrow~\pi\pi, K K$ to one such breaking term consistently with any other constraint.
Independently from this, both Minimal and Generic $U(2)^3$ are unlikely to give rise to any sizeable effect neither in top FCNCs nor in CP violation in $D$-$\bar{D}$ mixing at forseen experiments.
 
Suppose that a significant deviation from the SM emerged in the experiments to come, which could be accounted for in the effective framework described above. How could one tell that $U(2)^3$ is the relevant approximate symmetry, without uncovering by direct production the underlying dynamics (supersymmetry, a new strong interaction or whatever)? The best way would be to study in $B$ decays the correlation between the $s$ and the $d$ quarks in the final state, which would have to be the same as in the SM.
 Such correlation is in fact also expected in MFV. However, the only way to have effects in MFV similar to the ones discussed above in Minimal $U(2)^3$ requires the presence of two Higgs doublets, one coupled to the up quarks and one to the down quarks, with large values of the usual $\tan{\beta}$ parameter \cite{D'Ambrosio:2002ex,Kagan:2009bn}. This, in turn, would have other characteristic effects  not necessarily expected in Minimal $U(2)^3$. 
 In the case of small $\tan\beta$ or with one Higgs doublet only, distinguishing MFV from $U(2)^3$ would be straightforward by means of the additional effects discussed in \cite{Barbieri:2012uh}, like CP violation in $B_s$ mixing or non-universal contributions to $B\to K\nu\bar\nu$ vs.\ $K\to\pi\nu\bar\nu$ decays. A synthetic description of new physics effects in Minimal $U(2)^3$ and in Generic $U(2)^3$ is given in Table \ref{tab:u2u3}
 and compared with MFV (i.e. $U(3)^3$ at moderate or large $\tan{\beta}$). On top of the qualitative differences shown in Table \ref{tab:u2u3}, the size of the possible effects is significantly more constrained in $U(3)^3$ at moderate  $\tan{\beta}$ than in all other cases.
 
\section*{Acknowledgements}
This work was supported by the EU ITN ``Unification in the LHC Era'', 
contract PITN-GA-2009-237920 (UNILHC) and by MIUR under contract 2008XM9HLM.

\appendix

\section{Quark bilinears and effective operators}\label{app:bilinears}
\subsection{Diagonalization of the quark masses}

The effective operators of \eqref{eq:genL} are constructed from the most generic quark bilinears which contain the spurions and are formally invariant under $U(2)^3$.

To a sufficient approximation, the chirality conserving bilinears take the form
\begin{align}\label{cc}
\bar q_{Li}\gamma^{\mu}(X_L^{\alpha})_{ij} q_{Lj} &= a_L^{\alpha}\bar q_{3L}\gamma^{\mu}q_{3L} + b_L^{\alpha}\qLbar\gamma^{\mu}\qL + c_L^{\alpha}\bar q_{3L}\gamma^{\mu}\V^{\dag}\qL\notag\\
&+ d_L^{\alpha}(\qLbar \V)\gamma^{\mu}(\V^{\dag}\qL) + {\rm h.c.},\\
\bar u_{Ri}\gamma^{\mu}(X_{uR}^{\alpha})_{ij} u_{Rj} &= a_{uR}^{\alpha}\bar t_R\gamma^{\mu}t_R + b_{uR}^{\alpha}\uRbar\gamma^{\mu}\uR + c_{uR}^{\alpha}\bar t_R\gamma^{\mu}\Vu^{\dag}\uR\notag\\
&+ d_{uR}^{\alpha}(\uRbar \Vu)\gamma^{\mu}(\Vu^{\dag}\uR) + {\rm h.c.}\label{ccR},
\end{align}
where an analogous expression holds for the right-handed down quarks, we denote by uppercase letters the light generation doublets $\qL, \uR, \dR$, and all the parameters except the c's are real by hermiticity. These bilinears give rise to the four-fermion operators $\Delta\mathcal{L}^{4f}_{L,R}$, to $\Delta\mathcal{L}_{LR}^{4f}$, as well as to the kinetic terms.

Analogously, the chirality breaking bilinears are, to lowest order in the spurions,
\begin{align}
\bar q_{Li}(M_u^{\beta})_{ij} u_{Rj} &= \lambda_t\Big(a_u^{\beta}\bar q_{3L}t_R + b_u^{\beta}(\qLbar\V)t_R + c_u^{\beta}\qLbar\Delta Y_u \uR + d_u^{\beta}\bar q_{3L}(\Vu^{\dag}\uR)\nonumber\\
&+ e_u^{\beta}(\qLbar\V)(\Vu^{\dag}\uR)\Big) + {\rm h.c.}\, , \label{cb_u}
\end{align}
with an analogous expression for the down-quark sector and where all the parameters are complex. They generate the interaction terms $\Delta\mathcal{L}^{mag}$, $\Delta\mathcal{L}_{LR}^{4f}$, as well as the Yukawa couplings $Y_u, Y_d$.

Consider now the basis where the spurions take the form \eqref{genericV}, \eqref{genericY}. Moreover, notice that all the parameters in the kinetic and Yukawa terms, except one, can be made real through rephasings of the fields \cite{Barbieri:2012uh}. With these redefinitions, the previous operators can be written in the form
\begin{align}\label{ccmatrix}
X_{Lu}^{\rm kin} &= A_{uL}\mathbbm{1} + B_{uL} L_{23}^u\I_{32}^L (L_{23}^u)^{T}, & X_{Lu}^{\rm int,\alpha} &= A_{uL}^{\alpha}\mathbbm{1} + B_{uL}^{\alpha}U_{23}^{u,\alpha}\I_{32}^L (U_{23}^{u,\alpha})^{\dag},\\
X_{Ru}^{\rm kin} &= A_{uR}\mathbbm{1} + B_{uR} (R_{23}^u)^T\I_{32}^{Ru} R_{23}^u, & X_{Ru}^{\rm int,\alpha} &= A_{uR}^{\alpha}\mathbbm{1} + B_{uR}^{\alpha}(U_{23}^{u,\alpha})^{\dag}\I_{32}^{Ru} U_{23}^{u,\alpha},
\end{align}
plus analogous expressions for the down sector, where $\I_{32}^{I} = {\rm diag}(0, O(\epsilon^2_{I}), 1)$, the $A$'s and $B$'s are real functions of the parameters of \eqref{cc}, \eqref{ccR};
\begin{align}
Y_u &= \lambda_t (L_{23}^u\I_3 R_{23}^u + L_{12}^u\Delta \tilde Y^{\rm diag}_u V_{12}^u), & M_u^{\beta} &= \lambda_t(a_u^{\beta}{U}_{23}^{u,\beta}\I_3{V}_{23}^{u,\beta} + d_u^{\beta} L_{12}^u\Delta \tilde Y^{\rm diag}_u V_{12}^u),\\
Y_d &= \lambda_b (U_{23}^d\I_3 R_{23}^d + U_{12}^d\Delta \tilde Y^{\rm diag}_d V_{12}^d), & M_d^{\beta} &= \lambda_b(a_d^{\beta}{U}_{23}^{d,\beta}\I_3{V}_{23}^{d,\beta} + d_d^{\beta} U_{12}^d\Delta \tilde Y^{\rm diag}_d V_{12}^d),
\end{align}
where $\I_3 = {\rm diag}(0,0,1)$, $\Delta \tilde Y_{u,d}^{\rm diag} = {\rm diag} (y_{u,d}, y_{c,s},0)$, and $y_{u,d,c,s}$ are the diagonal entries of $\Delta Y_{u,d}^{\rm diag}$. Here and in the following $U_{ij}$ ($V_{ij}$) stand always for unitary left (right) matrices in the $(i,j)$ sector, while $L_{ij}$ ($R_{ij}$) indicate orthogonal left (right) matrices. In particular, in the notations of Section~\ref{sec:spurions}, $U_{12}^d = \Phi_L L_{12}^d$ and $V_{12}^{u,d} = \Phi_R^{u,d} R_{12}^{u,d}$.

We want to derive the expressions for these operators in the physical basis where the quark masses are diagonal, and the kinetic terms are canonical.
The kinetic terms are put in the canonical form by real rotations in the $(2,3)$ sector plus wavefunction renormalizations of the fields. One can check that these transformations do not alter, to a sufficient accuracy, the structure of the other operators, but cause only $O(1)$ redefinitions of the parameters.

The mass terms are diagonalized approximately by the transformation
\begin{align}
Y_u\mapsto Y_u^{\text{diag}} &= (L_{12}^u)^T(L_{23}^u)^T Y_u R_{23}^u R_{12}^u \equiv (L^u)^T Y_u R^u,\\
Y_d\mapsto Y_d^{\text{diag}} &=(U_{12}^d)^{\dag}(U_{23}^d)^{\dag} Y_d R_{23}^d V_{12}^d \equiv (U^d)^{\dag} Y_d V^d,
\end{align}
up to transformations of order $\epsilon_L y_{u,d,c,s}$, $\epsilon^u_R y_{u,c}$ and $\epsilon^d_R y_{d,s}$. Therefore one goes to the physical basis for the quarks by
\begin{align}
u_L&\mapsto L^u u_L, & d_L&\mapsto U^d d_L & u_R&\mapsto R^u u_R & d_R&\mapsto V^d d_R,
\end{align}
and the Cabibbo-Kobayashi-Maskawa matrix is
\begin{equation}
V_{CKM} \simeq (R_{12}^u)^T(R_{23}^u)^TU_{23}^d U_{12}^d\equiv (R_{12}^u)^T U_{23}^{\epsilon} U_{12}^d,
\end{equation}
where $U_{23}^{\epsilon}$ is a unitary transformation of order $\epsilon_L$.

In the physical mass basis the chirality conserving operators become
\begin{align}
X_{dL, \rm int}^{\alpha}&\mapsto A_{dL}^{\alpha}\mathbbm{1} + B_{dL}^{\alpha}(U_{12}^d)^{\dag} U_{23}^{d,\alpha}\I_{32}^L(U_{23}^{d,\alpha})^{\dag}U_{12}^d,\\
X_{dR, \rm int}^{\alpha}&\mapsto A^{\alpha}_{dR}\mathbbm{1} + B^{\alpha}_{dR}(V_{12}^d)^{\dag} V_{23}^{d,\alpha}\I_{32}^{Rd}(V_{23}^{d,\alpha})^{\dag}V_{12}^d,
\end{align}
and the $\sigma_{\mu\nu}$-terms are
\begin{equation}
M_d^{\beta}\mapsto \lambda_b\big(a_d^{\beta}(U_{12}^d)^{\dag} U_{23}^{\beta}\I_3 V_{23}^{\beta} V_{12}^d + c_d^{\beta} \Delta\tilde Y_d^{\rm diag}\big),
\end{equation}
plus analogous expressions for the up sector.

\subsection{List of interaction bilinears}
The following results are obtained after rotation to the mass basis, and factorizing out explicitly all the phases, CKM matrix elements and quark masses. We define $\xi_{ij} = V_{ti}^*V_{tj}$, and $\zeta_{ij} = V_{ib}V_{jb}^*$. In chirality breaking bilinears $\alpha = \gamma(g)$ for (chromo)electric dipole operators, while $\alpha = cb$ for other generic interaction bilinears. All the parameters are real.

\subsubsection{Up quark sector}
Chirality conserving LL, RR currents:
\begin{align}
X_{12}^{uL} &= c_D\zeta_{uc}, & X_{12}^{uR} &= \tilde c_D e^{i(\phi_1^u - \phi_2^u)}\zeta_{uc}\frac{s^u_R}{s^u_L}\left(\frac{\epsilon^u_R}{\epsilon_L}\right)^2,\\
X_{13}^{uL} &= c_t e^{i\phi_t}\zeta_{ut}, & X_{13}^{uR} &= \tilde c_t e^{i(\tilde\phi_t + \phi_1^u)}\zeta_{ut}\frac{s^u_R}{s^u_L}\frac{\epsilon^u_R}{\epsilon_L},\\
X_{23}^{uL} &= c_t e^{i\phi_t}\zeta_{ct}, & X_{23}^{uR} &= \tilde c_t e^{i(\tilde\phi_t + \phi_2^u)}\zeta_{ct}\frac{\epsilon^u_R}{\epsilon_L}.
\end{align}
Flavour conserving dipole operators:
\begin{align}
M_{11}^u &= c_D^{\rm \alpha}e^{i(\phi_D^{\rm \alpha} - \phi_1^u)}\zeta_{uu}\frac{s^u_R}{s^u_L}\frac{\epsilon^u_R}{\epsilon_L}, & M_{22}^u &= c_D^{\rm \alpha}e^{i(\phi_D^{\rm \alpha} - \phi_2^u)}\zeta_{cc}\frac{\epsilon^u_R}{\epsilon_L}, & M_{33}^u &= a_t e^{i\alpha_t}.
\end{align}
Flavour changing, chirality breaking operators:
\begin{align}
M_{12}^u &= c_D^{\rm \alpha}e^{i(\phi_D^{\rm \alpha} - \phi_2^u)}\zeta_{uc}\frac{\epsilon^u_R}{\epsilon_L}, & M_{21}^u  &= c_D^{\rm \alpha}e^{i(\phi_D^{\rm \alpha} - \phi_1^u)}\zeta_{uc}^*\frac{s^u_R}{s^u_L}\frac{\epsilon^u_R}{\epsilon_L},\\
M_{13}^u &= c_t^{\rm \alpha}e^{i\phi_t^{\rm \alpha}}\zeta_{ut}, & M_{31}^u &= \tilde c_t^{\rm \alpha}e^{i(\tilde\phi_t^{\rm \alpha} - \phi_1^u)}\zeta_{ut}^*\frac{s^u_R}{s^u_L}\frac{\epsilon^u_R}{\epsilon_L},\\
M_{23}^u &= c_t^{\rm \alpha}e^{i\phi_t^{\rm \alpha}}\zeta_{ct}, & M_{32}^u &= \tilde c_t^{\rm \alpha}e^{i(\tilde\phi_t^{\rm \alpha} - \phi_2^u)}\zeta_{ct}^*\frac{\epsilon^u_R}{\epsilon_L}.
\end{align}

\subsubsection{Down quark sector}
Chirality conserving LL, RR currents:
\begin{align}
X_{12}^{dL} &= c_K\xi_{ds}, & X_{12}^{dR} &= \tilde c_K e^{i(\phi_1^d - \phi_2^d)}\xi_{ds}\frac{s^d_R}{s^d_L}\left(\frac{\epsilon^d_R}{\epsilon_L}\right)^2,\\
X_{13}^{dL} &= c_B e^{i\phi_B}\xi_{db}, & X_{13}^{dR} &= \tilde c_B e^{i(\tilde\phi_B + \phi_1^d)}\xi_{db}\frac{s^d_R}{s^d_L}\frac{\epsilon^d_R}{\epsilon_L},\\
X_{23}^{dL} &= c_B e^{i\phi_B}\xi_{sb}, & X_{23}^{dR} &= \tilde c_B e^{i(\tilde\phi_B + \phi_2^d)}\xi_{sb}\frac{\epsilon^d_R}{\epsilon_L}.
\end{align}
Flavour conserving dipole operators:
\begin{align}
M_{11}^d &= \lambda_b\,c_K^{\rm \alpha}e^{i(\phi_K^{\rm \alpha} - \phi_1^d)}\xi_{dd}\frac{s^d_R}{s^d_L}\frac{\epsilon^d_R}{\epsilon_L}, & M_{22}^d &= \lambda_b\,c_K^{\rm \alpha}e^{i(\phi_K^{\rm \alpha} - \phi_2^d)}\xi_{ss}\frac{\epsilon^d_R}{\epsilon_L}, & M_{33}^d &= \lambda_b\,a_b e^{i\alpha_b}.
\end{align}
Flavour changing, chirality breaking operators:
\begin{align}
M_{12}^d &= \lambda_b\,c_K^{\rm \alpha}e^{i(\phi_K^{\rm \alpha} - \phi_2^d)}\xi_{ds}\frac{\epsilon^d_R}{\epsilon_L}, & M_{21}^d  &= \lambda_b\,c_K^{\rm \alpha}e^{i(\phi_K^{\rm \alpha} - \phi_1^d)}\xi_{ds}^*\frac{s^d_R}{s^d_L}\frac{\epsilon^d_R}{\epsilon_L},\\
M_{13}^d &= \lambda_b\,c_B^{\rm \alpha}e^{i\phi_B^{\rm \alpha}}\xi_{db}, & M_{31}^d &= \lambda_b\,\tilde c_B^{\rm \alpha}e^{i(\tilde\phi_B^{\rm \alpha} - \phi_1^d)}\xi_{db}^*\frac{s^d_R}{s^d_L}\frac{\epsilon^d_R}{\epsilon_L},\\
M_{23}^d &= \lambda_b\,c_B^{\rm \alpha}e^{i\phi_B^{\rm \alpha}}\xi_{sb}, & M_{32}^d &= \lambda_b\,\tilde c_B^{\rm \alpha}e^{i(\tilde\phi_B^{\rm \alpha} - \phi_2^d)}\xi_{sb}^*\frac{\epsilon^d_R}{\epsilon_L}.
\end{align}

\bibliographystyle{My}
\bibliography{LMFV}

\end{document}